# Observation of Valley Zeeman and Quantum Hall Effects at Q Valley of Few-Layer Transition Metal Disulfides

Zefei Wu[1†], Shuigang Xu[1†], Huanhuan Lu[1†], Gui-Bin Liu[3], Armin Khamoshi[2], Tianyi Han[1], Yingying Wu[1], Jiangxiazi Lin[1], Gen Long[1], Yuheng He[1], Yuan Cai[1], Fan Zhang[2*], Ning Wang[1*]

[1]*Department of Physics and the Center for 1D/2D Quantum Materials, the Hong Kong University of Science and Technology, Hong Kong, China*

[2]*Department of Physics, University of Texas at Dallas, Richardson, TX 75080, USA*

[3]*School of Physics, Beijing Institute of Technology, Beijing 100081, China*

[†]These authors contributed equally to this paper.

[*]E-mail: phwang@ust.hk; Zhang@utdallas.edu

In few-layer (FL) transition metal dichalcogenides (TMDC), the conduction bands along the ΓK directions shift downward energetically in the presence of interlayer interactions, forming six Q valleys related by three-fold rotational symmetry and time reversal symmetry. In even-layers the extra inversion symmetry requires all states to be Kramers degenerate, whereas in odd-layers the intrinsic inversion asymmetry dictates the Q valleys to be spin-valley coupled. In this Letter, we report the transport characterization of prominent Shubnikov-de Hass (SdH) oscillations for the Q valley electrons in FL transition metal disulfide (TMDs), as well as the first quantum Hall effect (QHE) in TMDCs. Our devices exhibit ultrahigh field-effect mobilities (~16,000 $cm^2V^{-1}s^{-1}$ for FL $WS_2$ and ~10,500 $cm^2V^{-1}s^{-1}$ for FL $MoS_2$) at cryogenic temperatures. Universally in the SdH oscillations, we observe a valley Zeeman effect in all odd-layer TMD devices and a spin Zeeman effect in all even-layer TMD devices.

Strong spin-orbit couplings in monolayer transition metal dichalcogenides (TMDC)[1, 2, 3, 4] entangle the spin and valley degrees of freedom of the valence band states at K valleys. This gives rise to exciting electronic and excitonic properties, such as optical circular dichroism[5, 6, 7, 8], opto-valley Hall effect[9, 10], and optical valley Zeeman effect[5, 6, 7, 8]. Inversion symmetry breaking in monolayer TMDCs plays an important role in opening energy gaps[5, 10] at the energy range of visible lights at K valleys. Together with the substantial spin-orbital couplings, the broken inversion symmetry in monolayer TMDCs further splits the spin up and spin down bands at the valence band edge of K valleys, which is absent in any bilayer TMDC. The former feature leads to the optical circular dichroism[5, 6, 7, 8] and valley Hall effect[9, 10], the latter feature yields the observation of optical valley Zeeman effect[5, 6, 7, 8]. As we will show, these peculiar electronic properties can be extended from monolayer TMDCs to FL TMDCs, from K valleys to Q valleys[11], and from optical probe to transport detection. This possibility can be explicitly appreciated by comparing the quantum transport measurements in odd-layer and even-layer TMDs, in which the essential inversion symmetry is broken and respected, respectively. So far the observed electron mobility in most atomically thin TMDCs, however, is unexpectedly low[12, 13, 14, 15] and has consequently prohibited the exploration of the quantum transport properties related to these fascinating layer-dependent electronic characteristics and their spin-valley entangled QHEs[16, 17, 18, 19]. Impurity scattering and non-ohmic contacts are two major obstacles to fabricating high-mobility TMDC-based field-effect devices. Unfavorable electrical contacts suppress electron injection into the channel material and dominate the electronic characteristics of the TMDC devices. Improving the electrical contacts to atomically thin TMDCs has become a major challenge in this research field[20, 21, 22, 23].

**High-mobility TMD devices.**

To eliminate any impurity effects induced during device fabrication, we employ a polymer-free dry transfer technique[24, 25] in an inert environment of argon or nitrogen, as schematically demonstrated in Figs. 1a–d. Using the encapsulation of FL TMDs in hexagonal boron nitride (BN) sheets, we can achieve high-quality low-temperature ohmic contacts and ultrahigh field-effect mobilities in TMD channels. For example, Figs. 1e–f show the $I_{SD}$–$V_{SD}$ curve of a nine-layer (9L) MoS$_2$ device. The linear characteristic of this device is observed at both 300 and 2 K. The contact resistivity at $T = 2$ K is around 0.25 kΩ·μm (see details in Supplementary Information).

The high quality of our BN-TMD-BN heterostructures is reflected by their four-terminal FET mobilities $\mu_F = \frac{1}{C_g} \frac{d\sigma}{dV_g}$ measured at different temperatures, where $\sigma$ is the conductivity and $C_g$ is the gate capacitance (1.1–1.2 F cm$^{-2}$, as obtained from Hall measurements). The FET characteristics of 9L MoS$_2$ and 6L WS$_2$ are shown in Fig. 1g and Fig. S3d (in Supplementary Information), whose FET mobilities at room temperature are ∼ 50 cm$^2$V$^{-1}$s$^{-1}$ and ∼ 300 cm$^2$V$^{-1}$s$^{-1}$, respectively. At $T = 2$ K, our TMD devices show excellent performance with remarkably improved FET mobilities ($\mu_F$ ∼ 10,500 cm$^2$V$^{-1}$s$^{-1}$ for 9L MoS$_2$ and $\mu_F$ ∼ 16,000 cm$^2$V$^{-1}$s$^{-1}$ for 6L WS$_2$). The phonon scattering is suppressed, and the corresponding Hall mobilities ($\mu_H = \frac{\sigma}{n_H e}$) reach 6,700 and 8,000 cm$^2$V$^{-1}$s$^{-1}$, where $n_H$ is the carrier density obtained from the Hall measurement. Markedly, the mobility in our 6L WS$_2$ device (∼16,000 cm$^2$V$^{-1}$s$^{-1}$) is more than 30 times higher than the previously reported record for WS$_2$ (∼ 486 cm$^2$V$^{-1}$s$^{-1}$)[13]. Moreover, $\mu_F$ and $\mu_H$ show a similar trend and follow the power law $\mu \sim T^{-\gamma}$ at $T > 70$ K (Figs. 1h–i). The mobilities become obviously saturated below 70 K, as a result of impurity scattering. The fitted exponent γ ranges from 2.01 for 6L WS$_2$ to 2.17 for 9L MoS$_2$, which are rather different from the previously reported values of 0.55–1.7[26, 27] but are highly consistent with the most recently reported values (1.9–2.5) for high-mobility MoS$_2$ devices[23]. No obvious relationship is observed between γ and sample thickness.

**Quantum oscillations in odd-layer TMDs.**

In the representative 9L MoS$_2$ device, the SdH oscillations in the longitudinal resistance $R$ appear at perpendicular magnetic fields $B > 4$ T. This property is the hallmark of the high quality and homogeneity of our BN-MoS$_2$-BN devices (Fig. 2a). Pronounced SdH oscillations are observed at relatively high gate voltages, where $\mu_H$ is sufficiently high. Quantitatively, at the low magnetic field range, the SdH oscillations in the longitudinal resistance $R$ of a single sub-band in 2D electron gas can be described by the Lifshitz–Kosevich formula[28]:

$$\frac{\Delta R}{R} = -4e^{-\pi/\omega_c \tau_q} \frac{\lambda}{\sinh\lambda} \cos(\frac{2\pi E_F}{\hbar\omega_c})$$

where $\lambda = 2\pi^2 k_B T/\hbar\omega_c$, $\omega_c = eB/m^*$ is the cyclotron frequency, $E_F = 2\pi\hbar^2 n/g_v g_s m^*$ is the Fermi energy, $g_v$ is the valley degeneracy, $g_s$ is the spin degeneracy, $m^*$ is the cyclotron mass of carriers, $k_B$ is the Boltzmann's constant, $n$ is the charge carrier density, and $\tau_q$ is the single particle scattering time. In a 2D electron gas with quadratic dispersons, the SdH oscillations can display useful information about the quantization of LLs when plotted versus 1/$B$. Fig. 2b shows the plots of $\Delta R$ (i.e., the background has been subtracted from $R$) as a function of 1/$B$ at different gate voltages $V_g$. The equal spacing between SdH valley positions (Figs. 2b–d) implies the single band nature at the studied $V_g$. Thus, extracting further information using the Lifshitz–Kosevich formula is appropriate. In principle, the periodicity of SdH oscillations is $1/B_F = g/\Phi_0 n$, where $g = g_s \cdot g_v$ is the LL degeneracy and $\Phi_0 = h/e$ is the flux quantum. At relatively high fields, the best fit of $n$ versus $B_F/\Phi_0$ (Fig. 2c) yields g = 3.0 ± 0.1; the linear fit of the LL filling factors versus the SdH valley positions (Fig. 2d) yields a zero Berry phase. As Fig. 2b shows, the filling factors $\nu = 36, 42,$ and 48 are also clearly observed at relatively low fields with a gate voltage $V_g = 40$ V. The degeneracy of 6 arises from the degeneracy between the 3 Q and 3 Q' valleys because the spin degeneracy within each Q or Q' valley is already lifted by the broken inversion symmetry in an odd-layer TMD. At relatively high magnetic fields, an LL sextet can be lifted into two LL triplets caused by the valley Zeeman effect, which is similar to the K/K' valley Zeeman effects observed most recently using optical circular dichroism[5, 6, 7, 8]. The Lande factor $g_L$ can be roughly estimated using the formula $g_L \mu_B B_C = k T_C$, where $\mu_B$ is the Bohr magneton, and $B_C$ is the lowest field and $T_C$ is the highest temperature for our observation of the valley Zeeman effect. With a filling factor of 33 at $V_g$ = 40 V, the valley Zeeman splitting disappears at ~10 K, amounting to $g_L$~3.4, which is comparable to those reported for $WSe_2$ and $MoSe_2$.

The cyclotron mass of charge carriers in the 9L $MoS_2$ device is obtained by investigating the temperature dependence of $\Delta R$ oscillations (Fig. 2e). For a given $E_F$ ($V_g$ = 60 V or $n = 4.32 \times 10^{12}$ cm$^{-2}$) and a given $B$, the $\Delta R$ peak amplitudes (Fig. 2f) follow the Ando formula[29] $\Delta R(T) \propto \frac{4\lambda(T)}{\sinh\lambda(T)} e^{-\frac{\pi}{\omega_c \tau_q}}$. We obtain $m^* \approx 0.27 \pm 0.01$ $m_e$, which is smaller than the effective mass (~0.5 $m_e$) obtained by our density functional theory (DFT) calculations. The corresponding quantum scattering time is $\tau_q$=206 fs, which is much shorter than the transport scattering time $\tau_t = \frac{m^*}{R_0 e^2 n}$ =1100 fs, thereby demonstrating that long-range scattering is dominant in our $MoS_2$ sample. (See Fig. S4 in Supplementary Information.)

Remarkably, the ultra-high mobility achieved in these TMD samples even enables us to observe the QHEs. Fig. 2g shows the longitudinal resistance $R$ and Hall resistance $R_{xy}$ as a function of $B$ at 2K in a 3L $MoS_2$ device. Beyond 6 T, $R_{xy}$ exhibits at least three well-quantized plateaus ($\nu$ =

36, 39, and 42), and they match very well with the corresponding *R* valleys, which are the hallmark of QHEs. To the best of our knowledge, QHEs have not been previously observed in any TMD device. As in other TMD devices with an odd number of layers, the SdH oscillations in the 3L MoS$_2$ device clearly exhibit an LL degeneracy of 3, implying the valley Zeeman splitting.

**Quantum oscillations in even-layer TMDs.**

The SdH oscillations in the representative 6L WS$_2$ device emerge when *B* field is greater than 2.5T (Fig. 3a). Although the gate voltages applied (V$_g$ = 50, 60, 70 V) to the 6L WS$_2$ device are similar to those applied (V$_g$ = 40, 60, 70 V) to the 9L MoS$_2$ device, the period of SdH oscillations appears twice larger in the 6L WS$_2$ device (Fig. 3b). Given that the experimentally accessible carrier density is low, the Fermi energy crosses only the lowest spin-degenerate sub-band at the Q/Q' valleys in our calculated band structure of 6L WS$_2$ (Fig. 4b). The single sub-band nature is also evidenced by the unique period in the SdH oscillations (Fig. 3b). The linear fit of *n* versus $B_F/\Phi_0$ indicates a LL degeneracy of ~11.8 ± 0.1 (Fig. 4c). At a large field of 6.5 T, the secondary SdH valleys and doubling of the oscillation frequency are clearly visible because of the spin Zeeman splitting of LL duodectets into LL sextets. The disappearance of secondary SdH valleys at around 10 K further indicates that the Lande factor is $g_L$~2.2. Under similar experimental conditions, the presence (absence) of SdH valleys as a result of the complete filling of an LL duodectet and/or sextet (an LL triplet) are repeatedly observed, e.g., in a 10L WS$_2$ device (g = 12 in Fig. 3g) and in a 10L MoS$_2$ device (g = 6 in Fig. 3h). In contrast to the data on odd-layer devices, such as 3L and 9L MoS$_2$, the phenomenon of even-layer devices exhibiting doubled LL degeneracies is rather universal.

The cyclotron mass $m^*$ in 6L WS$_2$ is also investigated (Figs. 3e–3f). Based on the ΔR data plotted as a function of $B$ and $k_BT$ at $V_g$ = 70V (n = $3.75 \times 10^{12}$ cm$^{-2}$), we obtain $m^* \approx 0.20 \pm 0.01\, m_e$ (Fig. 3f). This indirect experimental value of $m^*$ is again smaller than the effective mass $\sim 0.5\, m_e$ obtained in our DFT calculations. Based on the Ando formula, we further obtain the quantum scattering time $\tau_q$=586 fs in the 6L WS2 device, which is smaller than the corresponding transport scattering time $\tau_t$ =1300 fs. (See Fig. S4 in Supplementary Information.)

**Spin-valley coupled Q valleys in FL TMDs.**

Figs. 4a–b show the calculated band structures of 3L MoS2 and 6L WS2, in which the minima of the conduction bands are not located at the K/K' points, but rather at the Q/Q' points, that is, between K(K') and Γ points, with quadratic sub-bands. As illustrated in Fig. 4c, 3 Q and 3 Q' valleys exist in the first Brillouin zone of the FL TMDs. The rotational symmetry dictates the threefold Q valley degeneracy. For even-layer TMDs, the Q and Q' valleys are further related by both time reversal and spatial inversion symmetries, which require Kramers degeneracy. Consider the low carrier density in our 6L WS2 device, the Fermi energy is about 2.9 meV above the valley edge and crosses only one spin-degenerate sub-band at each valley (see the inset of Fig. 4b). Thus, in the SdH oscillations, we observe a 12-fold LL degeneracy at low fields and 6-fold LL degeneracy at high fields caused by the spin Zeeman splitting within each valley, that is, between |Q↑> and |Q↓> states (see Fig. 4d). The valley Zeeman effect is absent because of the inversion symmetry.

In contrast to the even-layer case, the inversion symmetry in the odd-layer devices is intrinsically broken; thus, all the sub-bands at each valley are spin non-degenerate. In the 3L MoS2 device (Fig. 4a), for instance, the Fermi energy is about 6.0 meV above the valley edge and crosses only the first sub-band, for which the spin up and down sub-bands are lifted by 4.3 meV. Thus, the SdH oscillations exhibit 6-fold LL degeneracy, which is reduced to 3-fold when the Zeeman effect is large, as shown Fig. 4d. The Zeeman effect in this case is obviously a valley Zeeman splitting between |Q↑> and |Q'↓> states, as a linear combination of the spin, orbital, and lattice Zeeman effects. Despite the complex orbital hybridizations and the strong spin-orbital couplings, because of time reversal symmetry orbital and spin characters are opposite for the Q and Q' valleys, which can be split in the presence of B field. The lattice Zeeman effect arises from the opposite Berry curvatures of Bloch electrons at two valleys, which is also dictated by time reversal symmetry. In the inversion symmetric even-layer cases, the Berry curvature vanishes and so does the lattice Zeeman effect. Nevertheless, the Q/Q' valley Zeeman effect, observed here for the first time in transport, is analogous to the K/K' valley Zeeman effect observed using optical circular dichroism[5, 6, 7, 8].

In summary, we demonstrate high-mobility FL TMD FETs achieved by encapsulating atomically thin TMDs between BN sheets. The QHE is clearly observed in our 3L-$MoS_2$ device for the first time in TMD transport. At moderate magnetic fields of 2.5–4 T and relatively low carrier density $\sim 10^{12}$ cm$^{-2}$, the quantum oscillations are dominated by the Q valleys, exhibiting a universal even–odd layer dependence. Above 4 T, we observe spin Zeeman effects in even-layer devices and valley Zeeman effects in odd-layer devices. The high-quality atomically thin BN-TMD-BN-based FETs fabricated in this work pave the way for understanding the multi-valley band structures of FL TMDs and for exploring their spin-valley entangled unconventional QHEs.

**References**

1.  Novoselov KS, Jiang D, Schedin F, Booth TJ, Khotkevich VV, Morozov SV, *et al.* Two-dimensional atomic crystals. *Proceedings of the National Academy of Sciences of the United States of America* 2005, **102**(30)**:** 10451-10453.

2.  Ayari A, Cobas E, Ogundadegbe O, Fuhrer MS. Realization and electrical characterization of ultrathin crystals of layered transition-metal dichalcogenides. *Journal of Applied Physics* 2007, **101**(1)**:** 014507.

3.  Mak KF, Lee C, Hone J, Shan J, Heinz TF. Atomically Thin $MoS_2$: A New Direct-Gap Semiconductor. *Physical Review Letters* 2010, **105**(13)**:** 136805.

4.  Wang QH, Kalantar-Zadeh K, Kis A, Coleman JN, Strano MS. Electronics and optoelectronics of two-dimensional transition metal dichalcogenides. *Nat Nano* 2012, **7**(11)**:** 699-712.

5.  Li Y, Ludwig J, Low T, Chernikov A, Cui X, Arefe G, *et al.* Valley Splitting and Polarization by the Zeeman Effect in Monolayer $MoSe_2$. *Physical Review Letters* 2014, **113**(26)**:** 266804.

6.  Srivastava A, Sidler M, Allain AV, Lembke DS, Kis A, Imamoglu A. Valley Zeeman effect in elementary optical excitations of monolayer $WSe_2$. *Nat Phys* 2015, **11**(2)**:** 141-147.

7.  Aivazian G, Gong Z, Jones AM, Chu R-L, Yan J, Mandrus DG, *et al.* Magnetic control of valley pseudospin in monolayer $WSe_2$. *Nat Phys* 2015, **11**(2)**:** 148-152.

8.  MacNeill D, Heikes C, Mak KF, Anderson Z, Kormányos A, Zólyomi V, *et al.* Breaking of Valley Degeneracy by Magnetic Field in Monolayer$MoSe_2$. *Physical Review Letters* 2015, **114**(3)**:** 037401.

9.  Mak KF, McGill KL, Park J, McEuen PL. The valley Hall effect in $MoS_2$ transistors. *Science*

2014, **344**(6191)**:** 1489-1492.

10. Xiao D, Liu G-B, Feng W, Xu X, Yao W. Coupled Spin and Valley Physics in Monolayers of MoS$_2$ and Other Group-VI Dichalcogenides. *Physical Review Letters* 2012, **108**(19)**:** 196802.

11. Liu H, Chen J, Yu H, Yang F, Jiao L, Liu G-B*, et al.* Observation of intervalley quantum interference in epitaxial monolayer WSe$_2$. *arXiv preprint arXiv:150707637* 2015.

12. Yu Z, Pan Y, Shen Y, Wang Z, Ong Z-Y, Xu T*, et al.* Towards intrinsic charge transport in monolayer molybdenum disulfide by defect and interface engineering. *Nat Commun* 2014, **5:** 5290.

13. Iqbal MW, Iqbal MZ, Khan MF, Shehzad MA, Seo Y, Park JH*, et al.* High-mobility and air-stable single-layer WS$_2$ field-effect transistors sandwiched between chemical vapor deposition-grown hexagonal BN films. *Scientific Reports* 2015, **5:** 10699.

14. Wang JIJ, Yang Y, Chen Y-A, Watanabe K, Taniguchi T, Churchill HOH*, et al.* Electronic Transport of Encapsulated Graphene and WSe$_2$ Devices Fabricated by Pick-up of Prepatterned hBN. *Nano Letters* 2015, **15**(3)**:** 1898-1903.

15. Larentis S, Fallahazad B, Tutuc E. Field-effect transistors and intrinsic mobility in ultra-thin MoSe$_2$ layers. *Applied Physics Letters* 2012, **101**(22)**:** 223104.

16. Xiao D, Yao W, Niu Q. Valley-Contrasting Physics in Graphene: Magnetic Moment and Topological Transport. *Physical Review Letters* 2007, **99**(23)**:** 236809.

17. Behnia K. Condensed-matter physics: Polarized light boosts valleytronics. *Nat Nano* 2012, **7**(8)**:** 488-489.

18. Li X, Zhang F, Niu Q. Unconventional Quantum Hall Effect and Tunable Spin Hall Effect in Dirac Materials: Application to an Isolated MoS$_2$ Trilayer. *Physical Review Letters* 2013,

**110**(6)**:** 066803.

19. Rycerz A, Tworzydlo J, Beenakker CWJ. Valley filter and valley valve in graphene. *Nat Phys* 2007, **3**(3)**:** 172-175.

20. Das S, Chen H-Y, Penumatcha AV, Appenzeller J. High Performance Multilayer MoS$_2$ Transistors with Scandium Contacts. *Nano Letters* 2013, **13**(1)**:** 100-105.

21. Gong C, Colombo L, Wallace RM, Cho K. The Unusual Mechanism of Partial Fermi Level Pinning at Metal–MoS$_2$ Interfaces. *Nano Letters* 2014, **14**(4)**:** 1714-1720.

22. Kappera R, Voiry D, Yalcin SE, Branch B, Gupta G, Mohite AD*, et al.* Phase-engineered low-resistance contacts for ultrathin MoS$_2$ transistors. *Nat Mater* 2014, **13**(12)**:** 1128-1134.

23. Cui X, Lee G-H, Kim YD, Arefe G, Huang PY, Lee C-H*, et al.* Multi-terminal transport measurements of MoS$_2$ using a van der Waals heterostructure device platform. *Nat Nano* 2015, **10**(6)**:** 534-540.

24. Wu Z, Han Y, Lin J, Zhu W, He M, Xu S*, et al.* Detection of interlayer interaction in few-layer graphene. *Physical Review B* 2015, **92**(7)**:** 075408.

25. Castellanos-Gomez A, Buscema M, Molenaar R, Singh V, Janssen L, Zant HSJvd*, et al.* Deterministic transfer of two-dimensional materials by all-dry viscoelastic stamping. *2D Materials* 2014, **1**(1)**:** 011002.

26. Radisavljevic B, Kis A. Mobility engineering and a metal–insulator transition in monolayer MoS$_2$. *Nat Mater* 2013, **12**(9)**:** 815-820.

27. Baugher BWH, Churchill HOH, Yang Y, Jarillo-Herrero P. Intrinsic Electronic Transport Properties of High-Quality Monolayer and Bilayer MoS$_2$. *Nano Letters* 2013, **13**(9)**:** 4212-4216.

28. Chen X, Wu Y, Wu Z, Han Y, Xu S, Wang L, *et al.* High-quality sandwiched black phosphorus heterostructure and its quantum oscillations. *Nat Commun* 2015, **6:** 7315.

29. Mancoff FB, Zielinski LJ, Marcus CM, Campman K, Gossard AC. Shubnikov-de Haas oscillations in a two-dimensional electron gas in a spatially random magnetic field. *Physical Review B* 1996, **53**(12)**:** R7599-R7602.

**Acknowledgements**

The authors thank Ivana Wong for her assistance in the sample preparation process. Financial support from the Research Grants Council of Hong Kong (Project Nos. 16302215, HKU9/CRF/13G, 604112 and N_HKUST613/12) and technical support from the Raith–HKUST Nanotechnology Laboratory at MCPF are hereby acknowledged. F.Z. and A.K. are supported by UT Dallas research enhancement funds. F.Z. is grateful to the Kavli Institute for Theoretical Physics for their hospitality during the finalization of this work, which is supported in part by the National Science Foundation under Grant No. PHY11-25915. G.B.L is supported by the NSFC of China with Grant No. 11304014.

**Author contributions**

N.W. and Z.W. conceived the projects. Z.W., S.X., and H.L. conducted the experiments, including the crystal growth, sample fabrication, data collection, and analyses. N.W. is the principal investigator and coordinator of this project. F.Z. and G.L. provided the physical interpretation and conducted the band structure calculations. Z.W., F.Z., and N.W. wrote the manuscript. The other authors provided technical assistance in the sample preparation, data collection/analyses, and experimental setup.

**Figure captions**

**Figure 1 | BN-TMD-BN heterostructure device.** (**a**) The sandwiched TMD heterostructure. (**b**) The BN-TMD-BN heterostructure for selective etching. The etching window is marked by arrows. (**c,d**) Optical (**c**) and schematic image (**d**) of a BN-TMD-BN FET device with a Hall bar configuration. Scale bar: 10 μm. (**e,f**) Two-terminal $I_{SD}$-$V_{SD}$ characteristics of a representative MoS$_2$ device at room temperature (**e**) and at low temperature 2 K (**f**). Linear I–V behavior is observed in both cases. (**g**) Four-terminal conductance in the WS$_2$ device plotted as a function of the gate voltage at various temperatures. (**h,i**) Field effect mobilities and Hall mobilities of MoS$_2$ (**h**) and WS$_2$ (**i**) at various temperatures.

**Figure 2 | Quantum oscillations in odd-layer TMDs**. (**a**–**f**) Quantum oscillations in 9L MoS$_2$. (**a**) Resistance changes as a function of $B$ field at +40 V (orange line), +60 V (blue line) and +70 V (black line) gate voltages. The inset shows the sample image. (**b**) $\Delta R$ plotted as a function of $1/B$ field yields an oscillation period $1/B_F$, which decreases with increasing gate voltages. The filling factors are labeled for the oscillations valleys. (**c**) The total carrier density n obtained from the Hall measurements as a function of $B_F/\Phi_0$ (black dots) for different gate voltages. The best fit (red line) indicates a Landau level degeneracy of ~3.0±0.1. (**d**) Landau level filling factors as a function of $1/B$ at different gate voltages. The linear fit yields a zero berry phase. (**e**) $\Delta R$ plotted as a function of $B$ at $V_g = 60$ V (n = $4.32 \times 10^{12}\ cm^{-2}$) at different temperatures. (**f**) Data points are the measured oscillation amplitude versus temperature $T$ of the peaks at different $B$ fields. The lines are fitted using the Lifshitz–Kosevich formula. The inset shows the fitted cyclotron mass ($0.27 \pm 0.01\ m_e$) at

different *B* fields. (**g**) Quantum Hall effect of 3L MoS$_2$. Magnetoresistance resistance *R* (blue line) and Hall resistance *R*$_{xy}$ (orange line) as a function of *B* field at 2K. The quantum Hall effect is shown by at least three well-quantized plateaus in *R*$_{xy}$ at $\nu =$ 36, 39, and 42.

**Figure 3 | Quantum oscillations in even-layer TMD.** (**a–f**) Quantum oscillations in 6L WS$_2$. (**a**) Resistance changes as a function of *B* field at +50 V (orange line), +60 V (blue line), and +70 V (black line) gate voltages. The inset shows the sample image. (**b**) Δ*R* plotted as a function of 1/*B* field yields an oscillation period 1/*B*$_F$, which becomes smaller at higher gate voltages. The filling factors are labeled for the oscillation valleys. (**c**) The total carrier density n obtained from the Hall measurements as a function of *B*$_F$/Φ$_0$ (black dots) for different gate voltages. The best fit (red line) indicates a Landau level degeneracy of ~11.8±0.1. (**d**) Landau level filling factors as a function of 1/*B* for different gate voltages. The linear fit yields a zero berry phase. (**e**) Δ*R* plotted as a function of *B* at V$_g$=50 V (n = $3.75 \times 10^{12}\ cm^{-2}$) at different temperatures. The arrow shows the Landau levels with spin Zeeman splitting at high *B* fields. (**f**) Data points are the measured oscillation amplitude versus temperature *T* of the peaks at different *B* fields. The lines are fitted using the Lifshitz–Kosevich formula. The inset shows the fitted cyclotron mass (0.20 ± 0.01 *m*$_e$) at different *B* fields. (**g**) Quantum oscillations in 10L WS$_2$ show a Landau level degeneracy of 12. (**h**) Quantum oscillations in 10L MoS$_2$ show a Landau level degeneracy of 6.

**Figure 4 | Layer-dependent spin-valley coupled Q valleys in TMDs.** (**a**) Calculated band structure of 3L MoS$_2$. The bottom of conduction band is located at the Q (Q') valleys. At the edge of each Q (Q') valley, two spin–split bands exist. The spin up and spin down sub-bands are lifted by 4.3 meV.

(**b**) Calculated band structure of 6L WS$_2$. The energy bands are spin degenerated at the edge of each Q (Q') valley. These spin-valley coupled band edges are further illustrated in (**c**), where the red and blue colors denote the spin-down and spin-up bands, respectively. Q$_1$, Q$_2$, and Q$_3$ have the same spin, and Q$_1$', Q$_2$', and Q$_3$' are their time reversals. (**d**) Energy level diagrams show the valley Zeeman effects in odd-layer devices (top) and the spin Zeeman effect in even-layer devices (bottom).

**Figure 1**

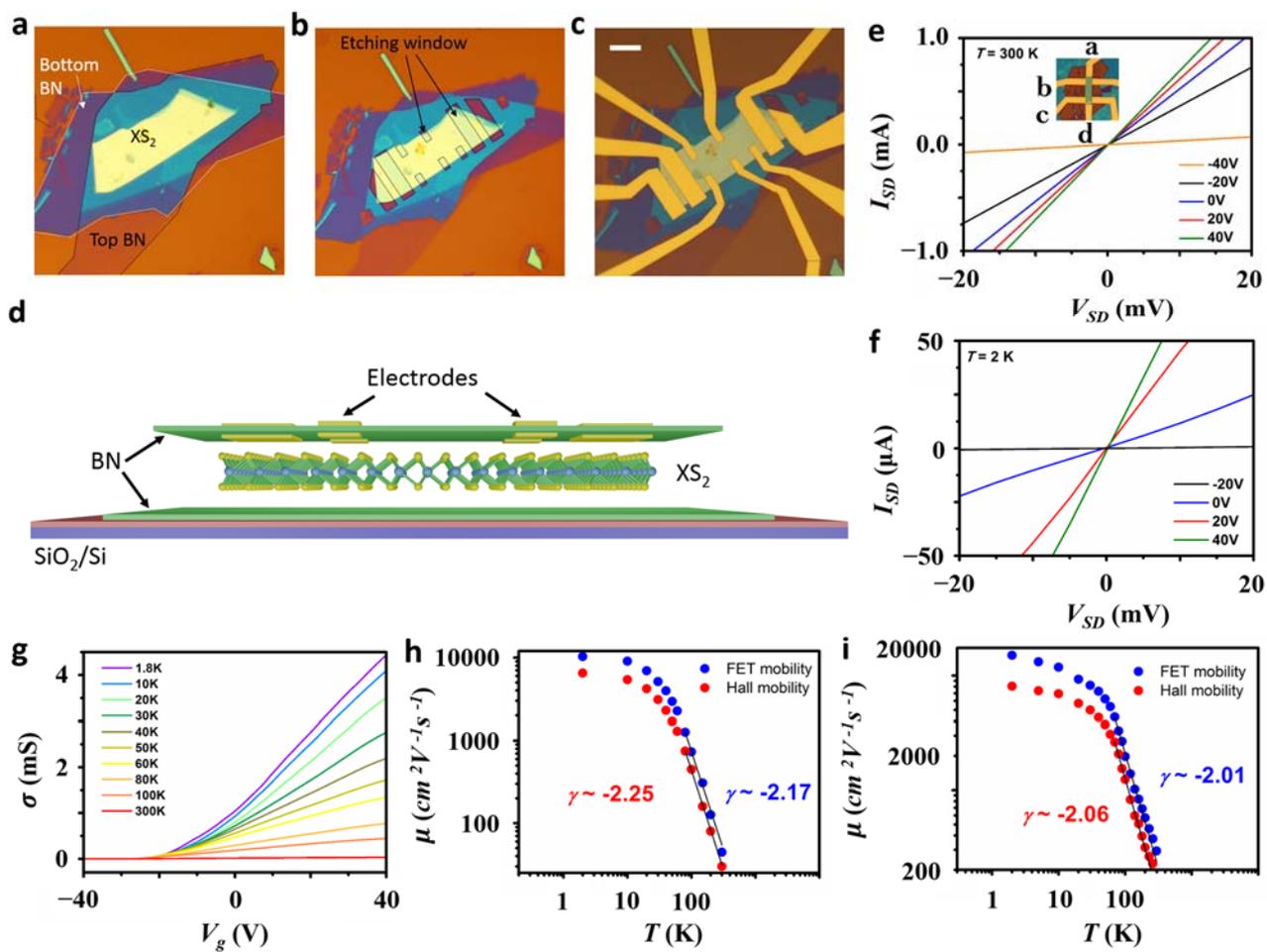

**Figure 2**

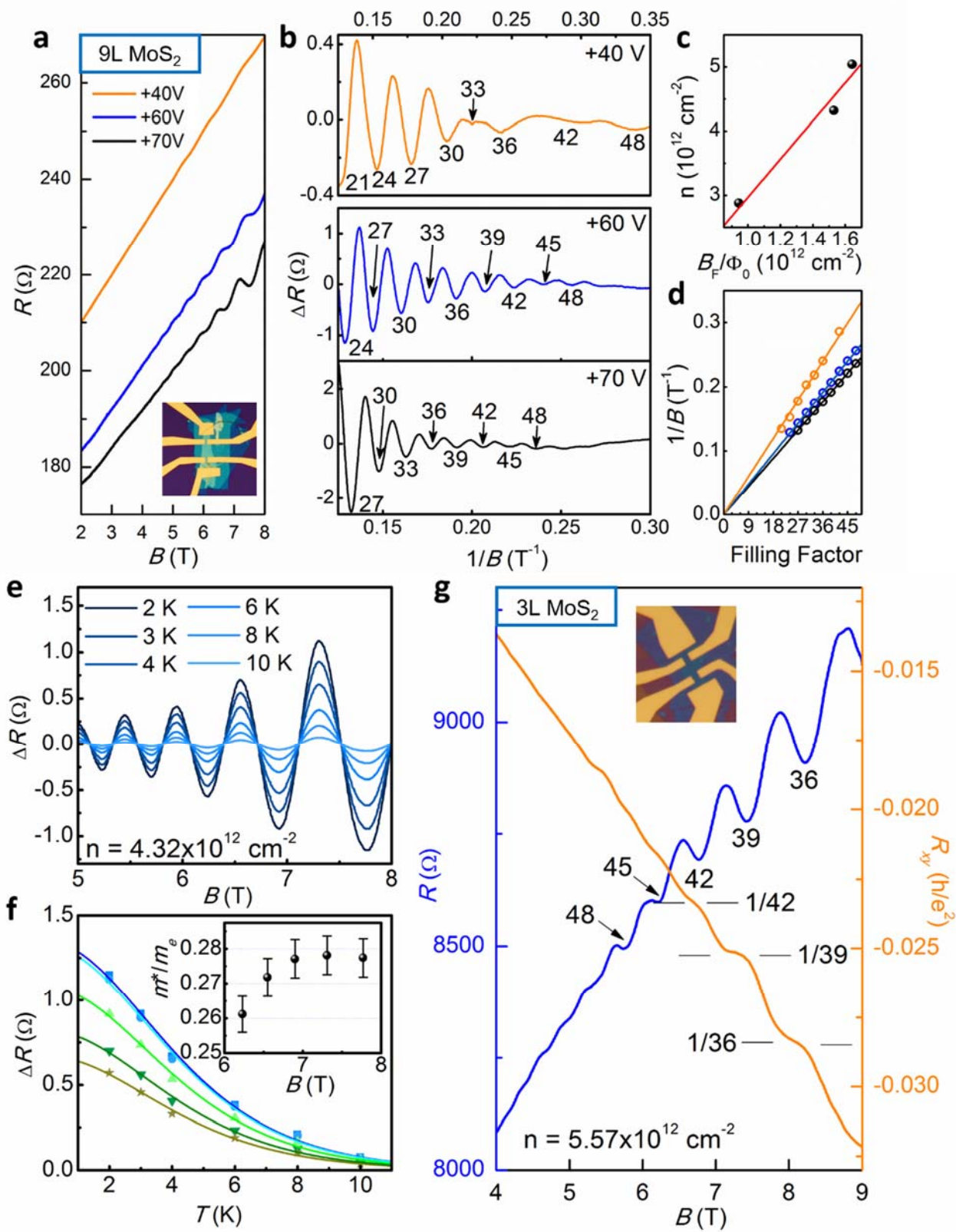

**Figure 3**

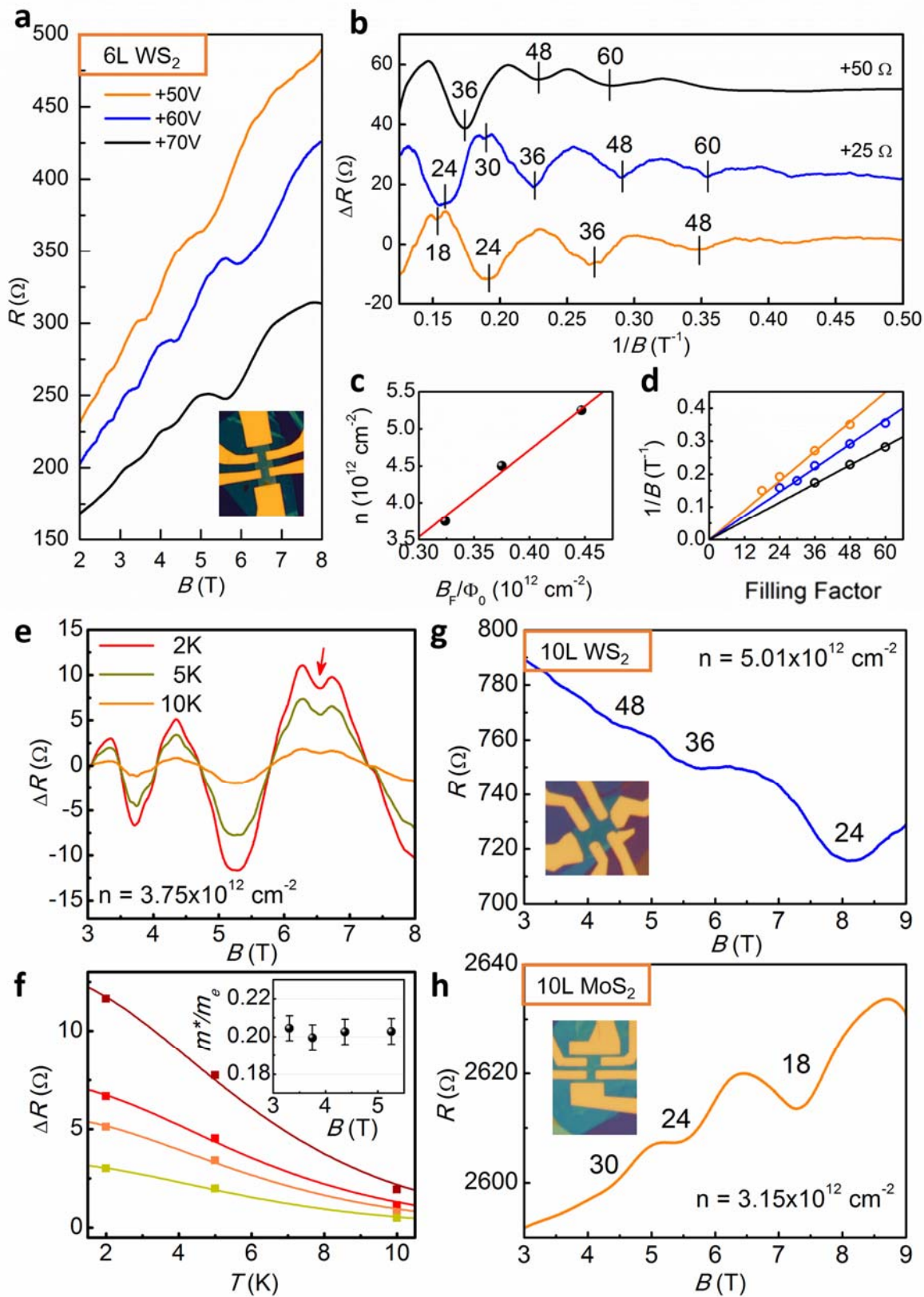

**Figure 4**

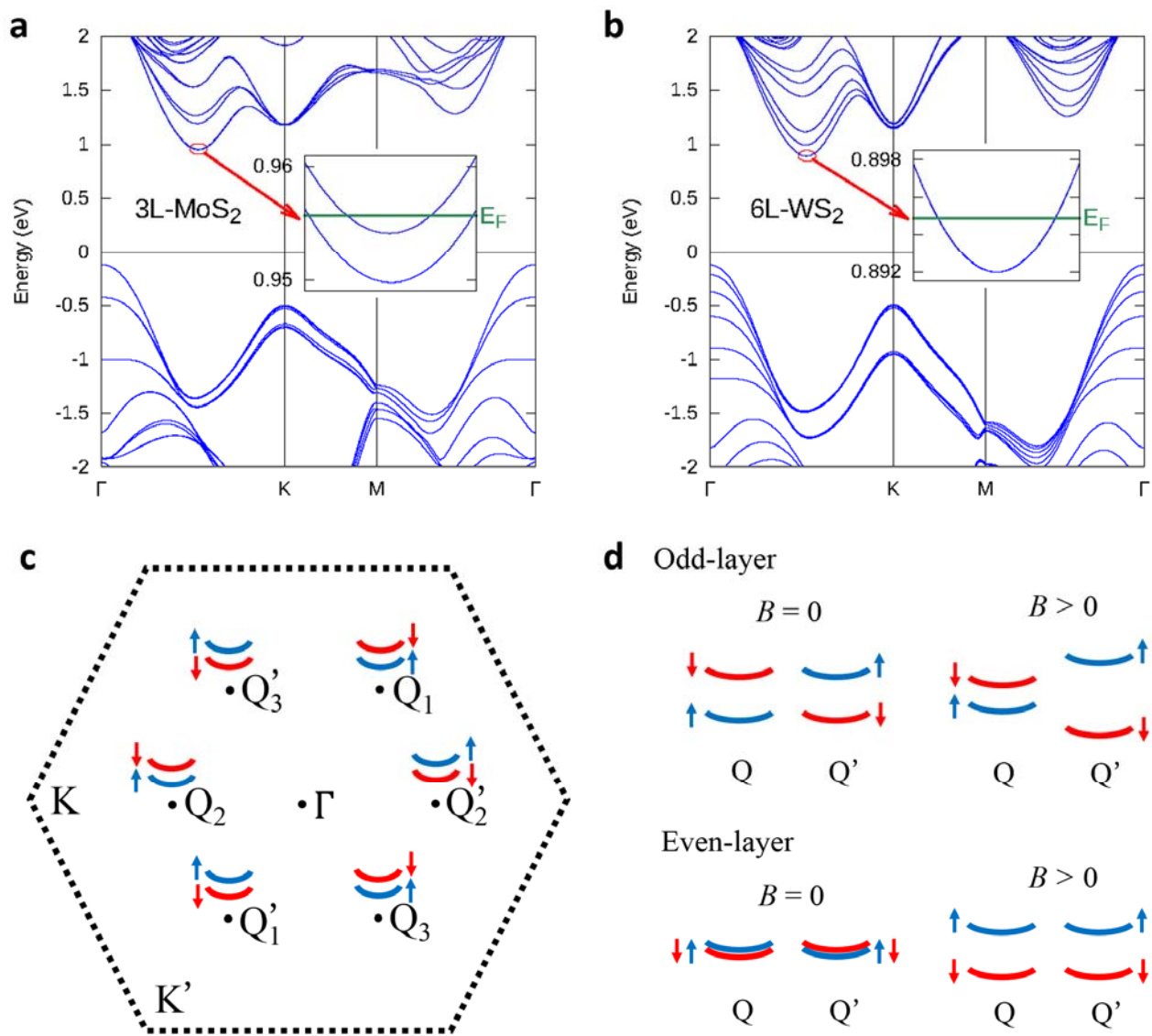

# Supplementary Information

## for

## Observation of Valley Zeeman and Quantum Hall Effects at Q Valley of Few-Layer Transition Metal Disulfides

Zefei Wu[1†], Shuigang Xu[1†], Huanhuan Lu[1†], Gui-Bin Liu[3], Armin Khamoshi[2], Tianyi Han[1], Yingying Wu[1], Jiangxiazi Lin[1], Gen Long[1], Yuheng He[1], Yuan Cai[1], Fan Zhang[2*], Ning Wang[1*]

[1]*Department of Physics and the Center for 1D/2D Quantum Materials, the Hong Kong University of Science and Technology, Hong Kong, China*

[2]*Department of Physics, University of Texas at Dallas, Richardson, TX 75080, USA*

[3]*School of Physics, Beijing Institute of Technology, Beijing 100081, China*

[†]These authors contributed equally to this paper.

[*]E-mail: phwang@ust.hk; Zhang@utdallas.edu

**1. Sample preparation and transport measurements.**

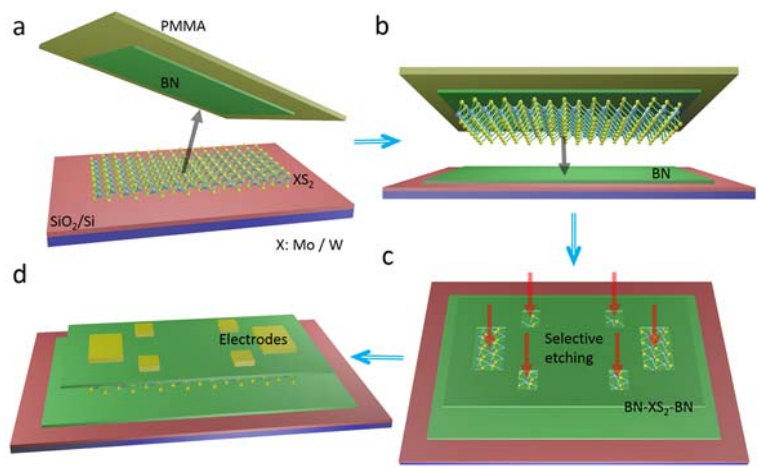

**Supplementary Fig. S1 | Fabrication of BN-XS$_2$-BN heterostructure device (X = Mo or W).** (**a**) A selected few-layer XS$_2$ is picked up from the SiO$_2$/Si substrate by a thin h-BN flake (5-15 nm thick) via van der Waals interactions. (**b**) The h-BN/XS$_2$ flake is then transferred onto a fresh thick h-BN flake (to eliminate possible influence of the SiO$_2$ substrate), which is exfoliated previously on a different SiO$_2$/Si substrate, to form a BN-XS$_2$-BN heterostructure. (**c**) The exposed top BN layer is then etched by oxygen plasma and XS$_2$ is partially exposed. (**d**) The electrodes are then patterned by e-beam lithography followed by standard e-beam

MoS$_2$ and WS$_2$ crystals are grown by the chemical vapor transport method reported previously[1]. The h-BN crystals (Polartherm grade PT110) are from Momentive. Atomically thin flakes are mechanically exfoliated on 300-nm SiO$_2$/Si substrates by the scotch-tape microcleavage method. To eliminate impurities induced during device fabrication, we employ the polymer-free dry transfer technique in an inert environment of argon or nitrogen, as schematically demonstrated in Fig. S1. The BN-XS$_2$-BN structure and the high-temperature annealing can guarantee the stability of our XS$_2$ devices. A selected few-layer XS$_2$ is picked up from the SiO$_2$/Si substrate by a thin h-BN flake (5-15 nm thick) via van der Waals interactions. The h-BN/XS$_2$ flake is then transferred onto a fresh thick h-BN flake (to eliminate possible influence of the SiO$_2$ substrate), which is exfoliated previously on a different SiO$_2$/Si substrate, to form a BN-XS$_2$-BN heterostructure. The atomically thin few-layer XS$_2$ is completely encapsulated by the top and bottom h-BN layers. In the annealing process (conducted in Ar atmosphere at 600 ℃), the small bubbles formed at the interfaces between h-BN and XS$_2$ are efficiently removed. While the top BN layer prevents the few-layer XS$_2$ from decomposing above 300 ℃, the high-temperature annealing helps reduce the charge trap density in XS$_2$.

To fabricate the metal electrodes, a hard mask is patterned on the BN-XS$_2$-BN heterostructure by the standard e-beam lithography technique using PMMA 950 A5. Since the etching rate of XS$_2$ by oxygen plasma was lower than that of h-BN, the exposed top BN layer is then etched by oxygen plasma and XS$_2$ is partially exposed. The electrodes are then patterned by e-beam lithography followed by standard e-beam evaporation (Ti/Au). Figures 1b-1e show the schematic and the optical images of a typical BN-XS$_2$-BN device with Hall-bar configurations. After the metal electrode deposition, the contact resistance is further reduced by a post annealing treatment at 300℃ for 2 hours.

The two-terminal *I-V* curves are measured by Keithley 6430. Other transport measurements are carried out using the standard lock-in technique (SR 830 and Keithley 6221 as the current source) in a cryogenic system. The cryogenic system provides temperature ranging from 1.8 K to 300 K and magnetic fields up to 9 T.

## 2. Determine the number of layers.

The thicknesses of XS$_2$ flakes are characterized by atomic force microscopy (Veeco-Innva). The number of layers N is determined by solving the equation:

$$N \times a + (N-1) \times b = d,$$

where a is the layer thickness, b is layer spacing, $c = 2a + 2b$ is the lattice constant, and d is the total thickness, as illustrated in Fig. S2. The measured sample thicknesses d$_m$ and their numbers of layers are listed in Table. S1.[2, 3]

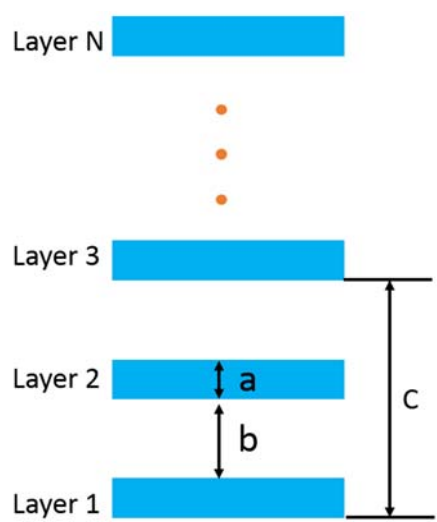

**Supplementary Fig. S2 | Schematic illustration of crystal structure.**

**Table. S1 | Sample thicknesses and the number of layers**

|  | a (nm) | b (nm) | d=Na+(N-1)b (nm) | $d_m$ (nm) |
|---|---|---|---|---|
| **9L-MoS$_2$** | 0.3172 | 0.2975 | 5.2348 | 5.37 |
| **3L-MoS$_2$** |  |  | 1.5466 | 1.62 |
| **10L-MoS$_2$** |  |  | 5.8495 | 5.90 |
| **6L-WS$_2$** | 0.3160 | 0.3015 | 3.4035 | 3.47 |
| **10L-WS$_2$** |  |  | 5.8735 | 5.90 |

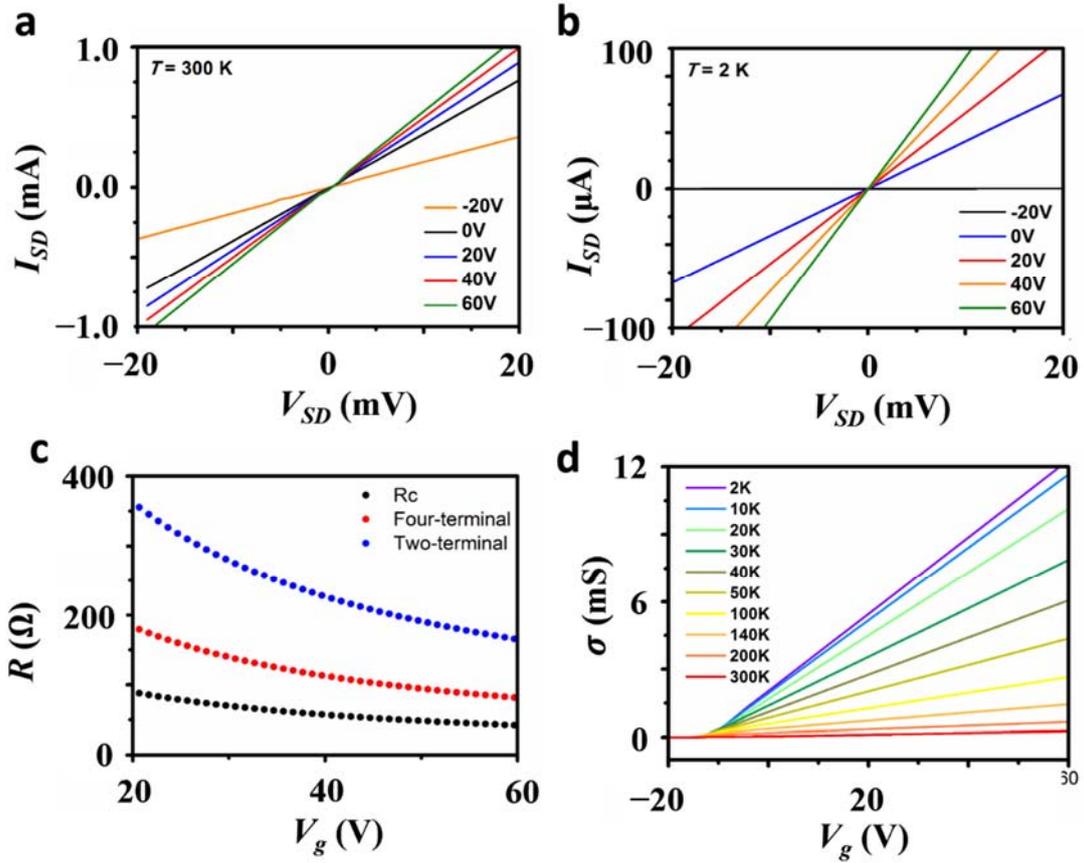

**Supplementary Fig. S3 | Output behavior and contact resistance of the 6L WS$_2$.** (**a,b**) Two-terminal ISD-VSD characteristics for the WS$_2$ device at room temperature (**a**) and at a low temperature of 2K (**b**). Linear behavior is observed in both cases. (**c**) Four-terminal, two-terminal, and calculated contact resistance $R_c$ for WS$_2$ at T=2 K. (**d**) Four-terminal conductance in the WS$_2$ device plotted as a function of the gate voltage at various temperatures.

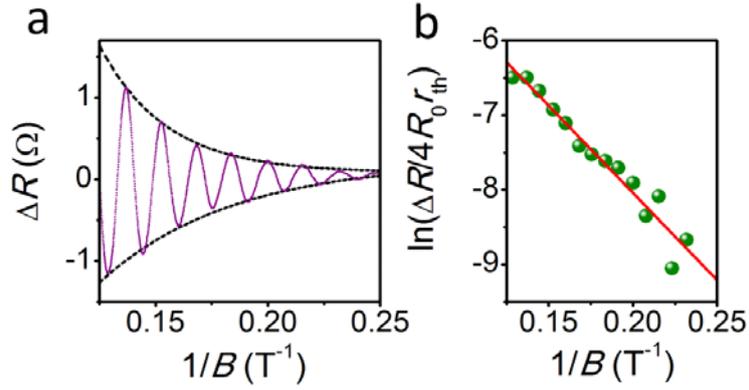

**Supplementary Fig. S4 | Quantum scattering time of the 9L-MoS$_2$.** (**a**) The solid line is Δ$R$ plotted as a function of inverse magnetic field B at V$_g$=60 V and $T$= 2 K. The dash lines are the amplitude fitted using the Ando formula. (**b**) Dingle plot of Δ$R$ for V$_g$=60 V and $T$= 2 K, and the extracted quantum scattering time is 206 ± 6 fs.

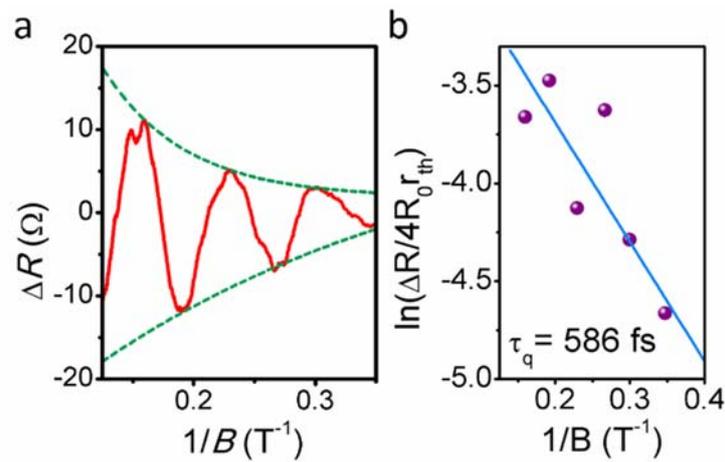

**Supplementary Fig. S5 | Quantum scattering time of the 6L-WS$_2$.** (**a**) The solid line is Δ$R$ plotted as a function of inverse magnetic field B at V$_g$=50 V and $T$= 2 K. The dash lines are the amplitude fitted using the Ando formula. (**b**) The Dingle plot of Δ$R$ for V$_g$=50 V and $T$= 2 K yields a quantum scattering time of 586 ± 68 fs.

## References

S1. Ubaldini A, Jacimovic J, Ubrig N, Giannini E. Chloride-Driven Chemical Vapor Transport Method for Crystal Growth of Transition Metal Dichalcogenides. *Crystal Growth & Design* 2013, **13**(10)**:** 4453-4459.

S2. Wildervanck JC, Jellinek F. Preparation and Crystallinity of Molybdenum and Tungsten Sulfides. *Zeitschrift für anorganische und allgemeine Chemie* 1964, **328**(5-6)**:** 309-318.

S3. Chang J, Register LF, Banerjee SK. Ballistic performance comparison of monolayer transition metal dichalcogenide MX2 (M = Mo, W; X = S, Se, Te) metal-oxide-semiconductor field effect transistors. *Journal of Applied Physics* 2014, **115**(8)**:** 084506.